\documentclass[aps,twocolumn,superscriptaddress,floatfix,longbibliography]{revtex4-1}
\usepackage{lipsum}
\usepackage{graphicx}
\usepackage[export]{adjustbox}
\usepackage{amsmath,amssymb,wasysym}
\usepackage{braket}
\usepackage{mathtools}
\usepackage{mathrsfs}
\usepackage{verbatim}
\usepackage{makecell}
\usepackage{cases}
\usepackage{bbm}
\usepackage{enumitem}
\usepackage[dvipsnames]{xcolor}
\usepackage{hyperref}

\newcommand{\im}{\mathbf{i}}

\newcommand{\Z}{\mathbb{Z}}
\newcommand{\C}{\mathbb{C}}

\DeclareMathOperator{\Tr}{Tr}

\newcommand{\Id}{\mathbb{I}}
\newcommand{\BigO}{\mathcal{O}}

\begin{document}
	\title{Origin and Limit of the Recovery of Damaged Information by Time Reversal}
	\author{Xiangyu Cao}
	 \affiliation{Laboratoire de Physique de l'Ecole Normale Sup\'erieure, ENS, Universit\'e PSL,
CNRS, Sorbonne Universit\'e, Universit\'e de Paris, 75005 Paris, France}
	\author{Thomas Scaffidi}
	\affiliation{Department of Physics, University of Toronto, Toronto, Ontario, M5S 1A7, Canada}
	
	\begin{abstract}
	   Recently it was found that scrambled information can be partially recovered by a time-reversed evolution, even after being damaged by an intruder. We reconsider the origin of the information recovery, and argue that the presence of classical chaos does not preclude it and only leads to a quantitative reduction of the recovery ratio. We also show how decoherence (i.e. entanglement with the intruder) limits the recovery, by proving an upper bound on the recovery ratio in terms of the entangling power of the intruder's action.
	\end{abstract}
	\maketitle

	\noindent\textit{Introduction}.-- Quantum dynamics scrambles local information by entangling many degrees of freedom. Although the scrambled information is no longer directly accessible, it is preserved in long-range correlations and can be recovered by applying the time-reversed unitary. In this sense, a scrambling unitary and its inverse can serve as an encoder-decoder. An intruder who attempts to access the encoded information by making a local measurement will not succeed to extract any useful information, but will create a perturbation which would be expected to disrupt the decoding process. It was shown recently~\cite{yan2020recovery}, however, that a finite amount of the encoded information can still be recovered after time-reversal.

	The physical origin of this finite recovery was presented in Ref.~\cite{yan2020recovery} as a consequence of the absence of classical chaos in quantum systems. The butterfly effect would indeed preclude any form of recovery due to the exponential amplification of the perturbation caused by the intruder during backwards time evolution. However, this interpretation leaves open the question of recovery in systems combining (semi-)classical and quantum degrees of freedom. In the first part of this work, we study the precise relation between recovery and chaos, and show in particular that recovery is still possible for a system combining quantum degrees of freedom with classical ones which exhibit classical chaos. We therefore propose that it is the finite dimensional Hilbert space of the target qudit hosting the initial information, rather than the absence of chaos, that is the physical origin of recovery.

	Another natural yet unaddressed question is how recovery is limited by the nature and strength of the perturbation performed by the intruder. Based on entanglement monogamy~\cite{coffman,horodecki09rev} and the fact that the scrambled information is stored nonlocally, one would expect recovery to get worse for perturbations which create more entanglement between the target qudit and the intruder's apparatus. In the second part of this work, we quantify this effect by deriving an upper bound on recovery in terms of the entangling power of the intruder's action~\cite{zanardi00entangling}.

\begin{figure}
        \centering
        \includegraphics[width=.8\columnwidth]{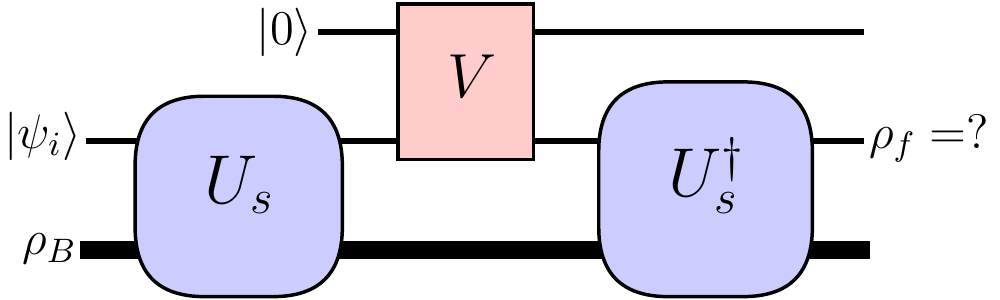}
        \caption{The information recovery process under consideration. The information $\psi_i$ stored in a qudit (middle) is encoded by a scrambler $U_s$ which couples the qudit to a bath (bottom). The information can be decoded by applying the time-reversed evolution, $U_s^{\dagger}$. What is the effect of an eavesdropping action (i.e. coupling with an ancilla qubit (top) by unitary operator $V$) ?    }
        \label{fig:process}
    \end{figure}

	Our analysis is based on the process shown in Fig.~\ref{fig:process} (our setup is slightly more general than \cite{yan2020recovery}). Alice, the encoder-decoder, prepares the \textbf{qudit} in a pure state $\rho_i = \vert \psi_i \rangle \langle \psi_i \vert$, and a \textbf{bath} in an arbitrary state $\rho_B$ (e.g., it can be the maximally mixed state); they are initially disentangled. We assume that the bath has a large Hilbert space.  She then applies a scrambling unitary $U_s$ on the qudit-bath system, encoding the information carried by $\rho_i$. Bob, the intruder, introduces an \textbf{ancilla} in a reference state $\vert 0 \rangle$ and couples it to the qudit with an eavesdropping unitary $V$.  Then, Alice decodes the (damaged) information by the time-reversal $U_s^{-1}$. We define the ratio of information recovery as  
    \begin{equation}
        r := \frac{\Tr[\rho_f \rho_i] - d^{-1}}{1 - d^{-1}} \,. \label{eq:def_r}
    \end{equation}
    where $\rho_f$ is the final reduced density matrix of the qudit, and $d$ is the dimension of the qudit Hilbert space. $r = 1$ corresponds to a perfect recovery $\rho_f = \rho_i$, and $r = 0$ if $\rho_f = \Id/d$ is the maximally mixed state with no information recovered. 

    
    \textit{Origin of Recovery.}    
    From the description of the protocol, it follows immediately that the overlap $\Tr[\rho_f \rho_i]$ which determines the recovery ratio is given by an out-of-time order correlator (OTOC)~\cite{larkin,Maldacena:2015waa}
	\begin{equation}
	    \Tr[\rho_i \rho_f] = 
	    \Tr[ \rho_i V(t) (\rho_i \otimes \rho_B) V(t)^{\dagger}] \,,\,  \label{eq:OTOC}
	\end{equation}
	where $V(t) = U_s^{-1} V U_s $,
	and the trace on the right hand side is over the total Hilbert space. Eq.~\eqref{eq:OTOC} is an example of ``fidelity OTOCs''~\cite{rey19fotoc}, which arise naturally in processes involving a time-reversal {(see \cite{yan2020loschmidt} and references therein for the relation with the Loschmidt echo)}.  In general, the recovery ratio depends on $\rho_i$, $\rho_B$, $U_s$, $V$, and can be only calculated numerically. Often, $U_s = e^{-\im t H}$ is generated by some Hamiltonian and one is interested in the time-dependence of the OTOC. Nevertheless, we can make analytical progress in two regimes where the OTOC is time independent:
	{
	\begin{enumerate}
	    \item the fully quantum scrambled regime, which is reached in the long time limit $t \to \infty$ for a generic (non-integrable) quantum Hamiltonian. We shall also take the limit of a large bath, but only after the long time limit.
	    \item A chaotic classical bath regime, where the bath admits a semiclassical description, and exhibits classical chaos. As we shall see below, This is a transient regime. 
	\end{enumerate}}
In both cases, $U_s$ can be approximated by a suitable random unitary. The (averaged) recovery ratio is then independent of the initial states, and is a function of $V$. We shall show that 
	 \begin{equation}
	r_{\text{q}} = \frac{f}{d^2} \,  \,,\, r_{\text{c}} =     \frac{f-1}{d^2 - 1} \,,  \label{eq:main} 
	\end{equation}
	where $f$ is defined as follows:
	\begin{equation}
	f := f[V] = \sum_{i,j=1}^d \sum_{a = 1}^{d_A}
	\langle j,0\vert V^\dagger \vert j,a\rangle  \langle i, a\vert V \vert i, 0\rangle   \,. \label{eq:def_f}
	\end{equation}
	Here $\{ \vert a \rangle, a = 1, \dots, d_A \}$ is a basis of the ancilla Hilbert space, and $\{ \vert i \rangle, i = 1, \dots, d\}$, that of the qudit; $\vert i, a \rangle := \vert i \rangle_{\text{qudit}} \vert a \rangle_{\text{ancilla}}$. Although $f$ is defined using a basis, one can check that it is basis independent. We can view it as an average fidelity, that measures how much $V$ preserves the input states~\footnote{We remark that $f$ is a function of the quantum channel which consists of acting with $V$ and tracing out the ancilla}.	
	
	The result~\eqref{eq:main}, which we will derive below, allows us to clarify the origin of the nonzero recovery. For concreteness let us consider the case where the intruder performs a strong measurement of the basis $\vert i \rangle$. That corresponds to the following control gate:
	\begin{equation} 
	V \vert i\rangle \vert 0 \rangle = \vert i\rangle \vert a_i \rangle  \,, \label{eq:strong_measurement}
	\end{equation}
	where $\vert a_1 \rangle, \dots, \vert a_d \rangle$ are orthonormal ancilla states. The general result~\eqref{eq:main} then implies
\begin{equation}
    	f = d \,,\, r_{\text{q}} = 1/d \,,\, r_{\text{c}} = 1/(d+1) \,.
\end{equation}
	The quantum result, in the qubit ($d = 2$) case, was found in \cite{yan2020recovery}. Putting it in a more general context, we see that the nonzero recovery ratio is not related to the absence of classical chaos, but simply due to the finite dimension of the qudit Hilbert space. The classical butterfly effect intuition mentioned above would apply if (and only if) the qudit itself becomes a classical degree of freedom itself, with $d \to \infty$; then we predict a vanishing recovery ratio, as expected. If the qudit remains quantum while the bath is classical and chaotic, the recovery ratio is still nonzero, albeit quantitatively lower.  
	
	We now derive the results, first in the full quantum scrambling regime. There, $U_s$ will resemble a random unitary, and a good approximation of the OTOC is obtained by averaging over $U(D)$ (with respect to the Haar measure), where $D$ is the  dimension of the qudit-bath Hilbert space. This can be evaluated using the following formula~\cite{collins06haar}:
	\begin{align}
	   & \overline{U_{i_1j_1} U_{k_1 \ell_1}^* U_{i_2j_2} U_{k_2 \ell_2}^* }  \\  =& \frac{1}{D^2 - 1} (\delta_{i_1k_1}\delta_{i_2k_2}
	   \delta_{j_1\ell_1}\delta_{j_2 \ell_2} + \delta_{i_1k_2}\delta_{i_2k_1}
	   \delta_{j_1\ell_2}\delta_{j_2 \ell_1}  ) \nonumber \\ & - 
	   \frac{1}{(D^2 - 1)D} (\delta_{i_1k_1}\delta_{i_2k_2}
	   \delta_{j_1\ell_2}\delta_{j_2 \ell_1} + \delta_{i_1k_2}\delta_{i_2k_1}
	   \delta_{j_1\ell_1}\delta_{j_2 \ell_2}  )  \nonumber
	\end{align}
  In the limit $D \to \infty$, the result is 
	\begin{equation}
	    \Tr[\rho_i \rho_f ] = 
	   \frac{f}{d^2} \left(1-\frac1d\right) + \frac1d + \BigO\left(\frac1D\right) \,,
	\end{equation}
	where $f$ is defined in \eqref{eq:def_f}. By \eqref{eq:def_r},  this is equivalent to $ r_{\text{q}} = {f}/{d^2} $ announced above \eqref{eq:main}.

    \begin{figure}
    	\centering
    	\includegraphics[width=.9\columnwidth]{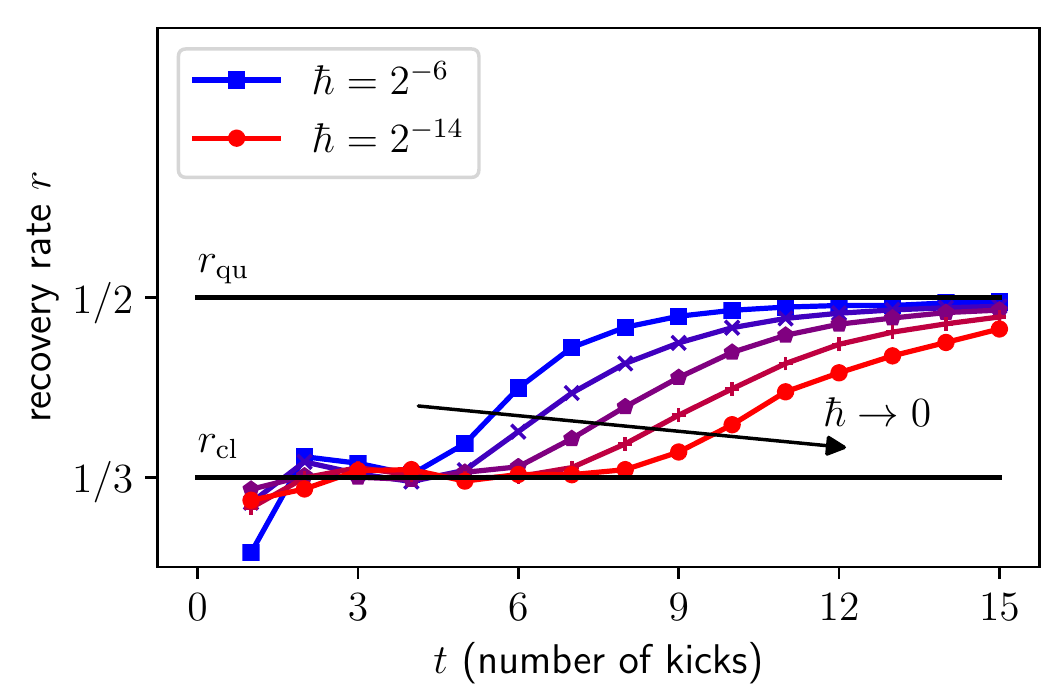}
    	\caption{Recovery ratio in a model of a qubit ($d = 2$) coupled to a kicked rotor~\eqref{eq:Hclassical} as a function of time, with $\hbar = 2^{-6}$, $2^{-8}$, \dots, $2^{-14}$ (from left to right). $V$ is a strong measurement \eqref{eq:strong_measurement}. In the semiclassical regime, we observe a crossover from the classical plateau $r = r_{\mathrm{cl}} = 1/3$ to the quantum plateau $r = r_{\mathrm{qu}} = 1/2$, at the Ehrenfest time scale $t_{\hbar} \propto \ln (1/\hbar)$. Other parameters are given by $K = 5$, $J = h = 1$.  The initial  state of the qudit is random on the Bloch sphere;  that of the bath is $\psi(p) \propto e^{-p^2/4}$. }
    	\label{fig:classical}
    \end{figure}
   The other saturating regime is when the qudit is coupled to a chaotic classical bath. Before deriving the recovery ratio~\eqref{eq:main}, let us illustrate the situation by a simple concrete example. Let the bath be a (quantum) kicked rotor~\cite{kicked} (see \cite{rozenbaum17,xu2020} on scrambling in this system), which is a single particle on a ring with a periodic coordinate $q \in [0, 2\pi)$, subject to the following Hamiltonian 
    \begin{equation}
    H_B(t) =  \frac{\hat{p}^2}2 +  K \sum_{n \in \Z}  \delta(t - n) \sin(\hat{q})
    \end{equation}
    where $\hat{q}$ and $\hat{p} =- \im\hbar \partial_q$ are position and momentum operators, respectively, and $K$ is the kicking strength. Classically, the phase space is completely chaotic when $K \gtrsim 1$. We now couple it to a qubit ($d=2$) by
    \begin{equation}
    H_{SB}(t) =  h {S}_z + J \sum_{n \in \Z} \delta(t - n) {S}_x \sin(\hat{q}) \,,
    \end{equation}
    where $h, J \ne 0$, and ${S}_\alpha = \frac\hbar2 \sigma_\alpha$ are  spin-half operators.  The total Hamiltonian is thus
    \begin{equation}
    H(t) = H_B(t) + H_{SB}(t) \,, \label{eq:Hclassical}
    \end{equation}  
    and we consider the resulting scrambling unitary {$U_s(t) = \mathrm{T} \exp\left(-\frac{\im}{\hbar} \int_0^t  H(s) d s \right)$} at after $t = 1, 2, 3, \dots$ kicks. We calculated the recovery ratio numerically as a function of $t$, see Fig.~\ref{fig:classical}. In the semiclassical ($\hbar \ll 1$) regime, we observe two distinct and well-established plateaus, connected by a crossover at the Erhenfest time $t_\hbar \sim \ln (1/\hbar)$. In semiclassical systems, $t_\hbar$ is the time after which an initial semiclassical wavepacket becomes too ``delocalized'' in the phase space due to classical chaos and loses its semiclassical description.
    Therefore, when $t \gg t_\hbar $, the classical nature of the bath becomes irrelevant and $r(t) \to r_{\text{q}}$ approaches the full quantum scrambling value. On the other hand, the intermediate plateau $1 \lesssim t \ll t_\hbar $ emerges within the time interval where bath admits a classical description and extends infinitely in the classical $\hbar \to 0$ limit. Like the quantum one, the classical plateau is robust: its value agrees with the analytical prediction \eqref{eq:main}, regardless of the parameters in the Hamiltonians and the initial state of the bath. 
    
    To understand quantitatively the classical plateau, we observe that the $H_{SB} \propto \hbar$, which, in the semiclassical regime, is much smaller than the bath Hamiltonian $H_B$. Thus, effectively, the bath is subject only to its own chaotic classical dynamics, while the qubit evolves with an effective single-body Hamiltonian \begin{equation}
    H_S^{\text{(eff)}}(t) = h {S}_z + J \sum_{n \in \Z} \delta(t - n) {S}_x \sin(q(t)) \label{eq:Heff}
    \end{equation} where $q(t) = \left<\hat{q}(t)\right>$ is a c-number provided by the classical evolution of the bath. Due to classical chaos, $q(t)$ is pseudo-random. Then, the time evolution operator generated by $H_S(t)$ becomes indistinguishable from a random unitary, but in $U(d)$.  The analytical prediction of $r_{\text{c}}$ \eqref{eq:main} results from an average over $U(d)$. {Curiously, we observed the classical plateau without classical chaos, e.g., in the Lipkin-Meshkov-Glick~\cite{LIPKIN1965188,*GLICK1965211,*MESHKOV1965199} model, if the bath is initially maximally mixed~\footnote{See Supplemental Material at [URL will be
 inserted by publisher], which contains Ref.~\cite{papparlardi18,xu2020}, for numerical results on the Lipkin-Meshkov-Glick model.}}
    
    To summarize, a chaotic classical bath effectively replaces the many-body quantum scrambling by a single-body ``scrambling''. This reduces the recovery ratio, but does not make it vanish. We note that $r = r_{\mathrm{c}}$ results from the bath being completely classical. Otherwise, we expect $r_{\mathrm{c}} \le r \le r_{\mathrm{q}}$, as observed during the crossover in the above example.  Note also that the qudit and the semiclassical bath are \textit{not} disentangled during the classical plateau; to the contrary, we observed numerically near-maximum entanglement (von Neumann entropy $\sim \ln d$), but the reduced density matrix of the bath still describes a well-localized classical configuration. This is possible because, for finite $d$, $d$ quantum states can correspond to a vanishing phase space volume $2\pi d \hbar \to 0$ in the semiclassical limit. 
    
    
    
  \noindent{\textit{Limit of Recovery.}}
   We now turn to a discussion of how the recovery ratio is limited by the nature and strength of the eavesdropping action. According to the above result~\eqref{eq:main}, this is entirely encoded in the quantity $f$. It is a trace, over the doubled Hilbert space $\C^d \otimes \C^d$ of the qudit, of the quantum channel induced by $V$ (upon tracing out the ancilla). As such, $f$ is not directly related to the entangling power of $V$ between the qudit and the ancilla. Nevertheless, as a second contribution of this Letter, we show that a large value $f \ge d$ or $r_{\mathrm{q}} = f/d^2 \ge 1/d$, implies a nontrivial bound on the entangling power~\cite{zanardi00entangling}, defined as the ancilla-qudit von Neumann entanglement entropy $\mathcal{S}$ after applying $V$ on a disentangled state $\vert k \rangle \vert 0 \rangle$, averaged over any basis $\{ \vert k \rangle \}_{k=1}^d$:
    \begin{figure}
        \centering
        \includegraphics[width=.8\columnwidth]{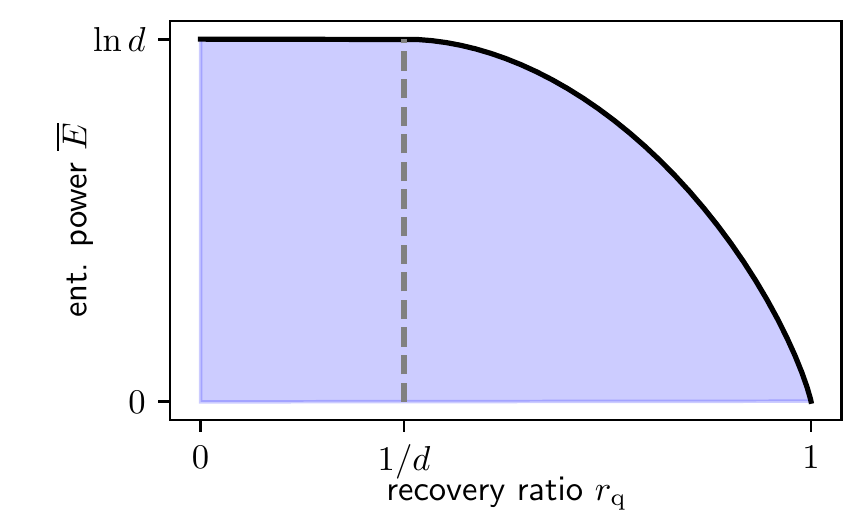}
        \caption{The bound~\eqref{eq:inequality_intro} constraining the average entangling power $\overline{E}$ of the eavesdropping gate coupling the qudit and the ancilla, and the recovery ratio $r_{\mathrm{q}}$ in the full quantum scrambling regime. Only the shaded region is allowed. Entanglement with the eavedropping ancilla (decoherence) limits the information recovery; with maximal entanglement ($\ln d$ entropy), the recovery ratio is $1/d$ at best. }
        \label{fig:Eandr}
    \end{figure}
     \begin{align}
      \overline{E} :=  \frac1d \sum_{k=1}^d \mathcal{S}[V \vert k, 0 \rangle ] \,. \label{eq:def_E}
    \end{align}
    We then claim that $\overline{E}$ and $r_{\text{q}}$ are constrained by the following bound (See Fig.~\ref{fig:Eandr} for a plot of the admissible region)
    \begin{equation}
          \overline{E} \le  -r_{\mathrm{q}} \ln r_{\mathrm{q}} - (1-r_{\mathrm{q}}) \ln \frac{1-r_{\mathrm{q}}}{d-1} \,,\, r_{\mathrm{q}} \ge 1/d \,. \label{eq:inequality_intro}
    \end{equation}
    This bound quantifies the the intuition that decoherence (caused by eavesdropping) limits the retrieval of scrambled information: a high recovery ratio guarantees the absence of decoherence. 
	

  We now prove \eqref{eq:inequality_intro}. Throughout the proof, we shall fix an arbitrary basis $\{ \vert i \rangle: i=1, \dots, d \}$ for the qudit and $\{ \vert a \rangle \}$ for the ancilla, and denote the matrix elements of $V$ in this basis as 
    \begin{equation}
        V_{ai, j} := \langle a, i \vert V \vert j, 0 \rangle \,,
    \end{equation}
    where we recall that $\vert 0\rangle$ is the reference ancilla state.

        The first step of the proof is a simple bound on $f$:
    \begin{align}
        f = &
        \sum_{a,i,j}  V_{a i,i} V_{a j,j}^* \le \sum_{a,i,j} 
         \frac12 ( |V_{a i,i}|^2 + 
     |V_{a j,j} |^2) \nonumber \\
      = & d \sum_{i,a} |V_{a i,i}|^2 \,. \label{eq:f_bound0}
    \end{align}
 By introducing 
 \begin{equation}
        \delta_i := \sum_a |V_{a i,i}|^2 \,, \label{eq:delta_i}
   \end{equation}
  and a shorthand for average over the basis:
   \begin{equation}
       \mathbb{E}_i [\dots] := \frac1d \sum_{i=1}^d [\dots] \,,
   \end{equation}  
 we rewrite \eqref{eq:f_bound0} as follows (recall also that $r_{\mathrm{q}} = f / d^2$):
 \begin{equation}
     r_{\mathrm{q}} \le \mathbb{E}_i \delta_i  \,. \label{eq:randdelta}
 \end{equation}

  Most of the remainder of the proof consists in relating $\delta_i$ with the entanglement entropy of  the state
  $$ \vert  V i \rangle := V \vert i,0 \rangle  \,,$$ for each $i$. It is not hard to show that $\delta_i$ is a diagonal matrix element of the reduced density matrix of the qudit:
  \begin{equation}
\rho_i := \Tr_{\text{ancilla}} \vert V i \rangle \langle V i \vert \Rightarrow  \langle i \vert \rho_i \vert i \rangle  = \delta_i  \,. \label{eq:matrix_ele}
\end{equation}
Therefore the largest eigenvalue value of $\rho_i$, denoted $s_{i,1}$, is at least $\delta_i$: 
\begin{equation}
    s_{i,1} \ge \delta_i \,. \label{eq:s_delta}
\end{equation}
Note that $s_{i,1}$ is the largest Schmidt value contributing to the entanglement entropy. When $\delta_i \ge 1/d$, the inequality \eqref{eq:s_delta} is nontrivial, and implies that the von Neumann entropy $\mathcal{S}$ cannot be larger than that of the Schmidt values:
$$ \delta_i, \underbrace{\frac{1-\delta_i}{d-1}, \dots, \frac{1-\delta_i}{d-1}}_{(d-1) \text{ times}} \,.  $$
When $\delta_i < 1/d$, \eqref{eq:s_delta} is trivial, but we have still the general bound $\mathcal{S}\le \ln d$. Combining the two cases, we obtain 
\begin{equation}
    \mathcal{S} [\vert V i \rangle] \le g(\delta_i) \,, \label{eq:bound_Si} 
\end{equation}
where the function $g$ is defined as 
\begin{equation}
    g(x \ge 1/d ) := 
       -x\ln x - (1-x) \ln \frac{1-x}{d-1} \,,
\end{equation}
and $g(x < 1/d) := \ln d$. One can check that $g$ is decreasing and concave, $g'(x) \le 0$, $g''(x) \le 0$ almost everywhere. Then we have
\begin{equation}
    \overline{E} = \mathbb{E}_i \mathcal{S} [\vert V i \rangle] \le \mathbb{E}_i g(\delta_i) \le 
    g( \mathbb{E}_i \delta_i) \le g(r_{\mathrm{q}}) \label{eq:proofover}
\end{equation}
where we used in turn the definition of $\overline{E}$~\eqref{eq:def_E}, \eqref{eq:bound_Si}, Jensen's inequality, and \eqref{eq:randdelta}. This completes the proof as \eqref{eq:proofover} is equivalent to \eqref{eq:inequality_intro} announced above. 

A few remarks are in order. First, it is straightforward to adapt the above argument to other entanglement measures. It suffices to modify the function $g$ accordingly. For example, for the average $n$-th Renyi purity, we have 
\begin{align*}
  &  \mathbb{E}_i \Tr[\rho_i^n] \ge g_2(r_{\mathrm{q}}) \,,\,   g_2(x \ge 1/d) :=  x^n + \frac{(1-x)^n}{(d-1)^{n-1}} \,,
\end{align*}
and $g_2(x < 1/d) = 1/d.$ Second, the basis dependence of the average might seem unappealing, but since the inequality holds for any basis, we can further average over all bases and obtain the same bound on the Hilbert sphere average. 
Finally, the above bound is tight, and is saturated by a family of ``weak measurement'' unitaries interpolating between a strong measurement \eqref{eq:strong_measurement} and no action. They are parametrized by $\epsilon \in [0,1]$ (measurement strength), and defined by
\begin{equation}
   V \vert i\rangle \vert 0 \rangle =  \vert i \rangle \vert a_i \rangle 
\end{equation}
where $\vert a_i \rangle, i = 1, \dots, d$ are non-orthogonal ancilla states satisfying
\begin{equation} 
    \left< a_i \vert a_j \right> = (1-\epsilon )+ \epsilon \delta_{ij} \,. 
\end{equation}
It follows that the recovery ratio is
\begin{equation}
    r_{\mathrm{q}} = \frac1{d^2}\sum_{ij}  \left< a_i \vert a_j \right> =  (1-\epsilon) + \frac{\epsilon}d \,.
\end{equation}

Now, the states $ V  \vert i \rangle \vert 0 \rangle $ are disentangled, but let us consider another basis of states:
\begin{equation}
    \vert k_X \rangle = \frac1{\sqrt{d}} \sum_{i}   \omega_{(ik)} \vert i \rangle \,,\, \omega_{m} := e^{2\pi\im m / d} \,.  \label{eq:basis_kX}
\end{equation}
An explicit calculation shows that the reduced density matrix of $  V \vert k_X \rangle \vert 0 \rangle$ to the qudit is
\begin{align}
    \rho_{k_X}
   &= (1-\epsilon) | k \rangle \langle k | + \epsilon \Id/d \,,
\end{align}
and that the averaged entanglement entropy over the basis \eqref{eq:basis_kX} saturates the bound~\eqref{eq:inequality_intro}. 

%
    
\noindent\textit{Discussion}-- This work addressed two questions on the recovery of damaged information by a time-reversal protocol~\cite{yan2020recovery}. We showed that recovery is not incompatible with classical chaos, and we proved a bound relating recovery and decoherence.

We proved an upper bound on the recovery ratio. Is there a lower bound? The naive answer is no, since $f$ can vanish even without ancilla entanglement (it suffices to choose $V\in U(d)$ with $\Tr[V] = 0$). Nevertheless, we can remove this trivial effect by adding a single-site gate $v$ acting on the qudit (without changing the entangling power) and by considering whether $\tilde{f} = \max_{v \in U(d)} f[v V]$ has a nontrivial lower bound as a function of the entangling power. Numerical studies in small Hilbert spaces indicate that this is the case, but {proving a lower bound seems non-trivial and is left for future work.}

Our analysis relied on approximating quantum scrambling by a single random unitary, reducing the calculation to a few-body one. However, this is only valid if the scrambling-rewind is perfectly carried out.
A natural extension of this work would be to consider the effect of imperfect time evolution, e.g. due to gate errors or decoherence, modeled by a ``hybrid'' quantum circuit (containing unitary and non-unitary gates), with forward and backward evolution. Such a setup would give rise to a double-folded Keldysh contour, which is necessary to detect the distinct entanglement phases that were recently found in hybrid quantum dynamics~\cite{li18transition,nahum18transition,bao19theory,jian19theory}. How these phases affect information recovery is an interesting question that we leave for future work.   

\acknowledgements{XC acknowledges support from a US Department of Energy grant DE-SC0019380 during his postdoctoral appointment at University of California, Berkeley, where this work was initiated. TS acknowledges the support of the Natural Sciences and Engineering Research Council of Canada (NSERC), in particular the Discovery Grant [RGPIN-2020-05842], the Accelerator Supplement [RGPAS-2020-00060], and the Discovery Launch Supplement [DGECR-2020-00222]. TS contributed to this work prior to joining Amazon.}
	\bibliography{ref}

\begin{thebibliography}{22}%
\makeatletter
\providecommand \@ifxundefined [1]{%
 \@ifx{#1\undefined}
}%
\providecommand \@ifnum [1]{%
 \ifnum #1\expandafter \@firstoftwo
 \else \expandafter \@secondoftwo
 \fi
}%
\providecommand \@ifx [1]{%
 \ifx #1\expandafter \@firstoftwo
 \else \expandafter \@secondoftwo
 \fi
}%
\providecommand \natexlab [1]{#1}%
\providecommand \enquote  [1]{``#1''}%
\providecommand \bibnamefont  [1]{#1}%
\providecommand \bibfnamefont [1]{#1}%
\providecommand \citenamefont [1]{#1}%
\providecommand \href@noop [0]{\@secondoftwo}%
\providecommand \href [0]{\begingroup \@sanitize@url \@href}%
\providecommand \@href[1]{\@@startlink{#1}\@@href}%
\providecommand \@@href[1]{\endgroup#1\@@endlink}%
\providecommand \@sanitize@url [0]{\catcode `\\12\catcode `\$12\catcode
  `\&12\catcode `\#12\catcode `\^12\catcode `\_12\catcode `\%12\relax}%
\providecommand \@@startlink[1]{}%
\providecommand \@@endlink[0]{}%
\providecommand \url  [0]{\begingroup\@sanitize@url \@url }%
\providecommand \@url [1]{\endgroup\@href {#1}{\urlprefix }}%
\providecommand \urlprefix  [0]{URL }%
\providecommand \Eprint [0]{\href }%
\providecommand \doibase [0]{http://dx.doi.org/}%
\providecommand \selectlanguage [0]{\@gobble}%
\providecommand \bibinfo  [0]{\@secondoftwo}%
\providecommand \bibfield  [0]{\@secondoftwo}%
\providecommand \translation [1]{[#1]}%
\providecommand \BibitemOpen [0]{}%
\providecommand \bibitemStop [0]{}%
\providecommand \bibitemNoStop [0]{.\EOS\space}%
\providecommand \EOS [0]{\spacefactor3000\relax}%
\providecommand \BibitemShut  [1]{\csname bibitem#1\endcsname}%
\let\auto@bib@innerbib\@empty
\bibitem [{\citenamefont {Yan}\ and\ \citenamefont
  {Sinitsyn}(2020)}]{yan2020recovery}%
  \BibitemOpen
  \bibfield  {author} {\bibinfo {author} {\bibfnamefont {Bin}\ \bibnamefont
  {Yan}}\ and\ \bibinfo {author} {\bibfnamefont {Nikolai~A.}\ \bibnamefont
  {Sinitsyn}},\ }\bibfield  {title} {\enquote {\bibinfo {title} {{Recovery of
  damaged information and the out-of-time-ordered correlators}},}\ }\href
  {\doibase 10.1103/PhysRevLett.125.040605} {\bibfield  {journal} {\bibinfo
  {journal} {Phys. Rev. Lett.}\ }\textbf {\bibinfo {volume} {125}},\ \bibinfo
  {pages} {040605} (\bibinfo {year} {2020})}\BibitemShut {NoStop}%
\bibitem [{\citenamefont {Coffman}\ \emph {et~al.}(2000)\citenamefont
  {Coffman}, \citenamefont {Kundu},\ and\ \citenamefont {Wootters}}]{coffman}%
  \BibitemOpen
  \bibfield  {author} {\bibinfo {author} {\bibfnamefont {Valerie}\ \bibnamefont
  {Coffman}}, \bibinfo {author} {\bibfnamefont {Joydip}\ \bibnamefont {Kundu}},
  \ and\ \bibinfo {author} {\bibfnamefont {William~K.}\ \bibnamefont
  {Wootters}},\ }\bibfield  {title} {\enquote {\bibinfo {title} {Distributed
  entanglement},}\ }\href {\doibase 10.1103/PhysRevA.61.052306} {\bibfield
  {journal} {\bibinfo  {journal} {Phys. Rev. A}\ }\textbf {\bibinfo {volume}
  {61}},\ \bibinfo {pages} {052306} (\bibinfo {year} {2000})}\BibitemShut
  {NoStop}%
\bibitem [{\citenamefont {Horodecki}\ \emph {et~al.}(2009)\citenamefont
  {Horodecki}, \citenamefont {Horodecki}, \citenamefont {Horodecki},\ and\
  \citenamefont {Horodecki}}]{horodecki09rev}%
  \BibitemOpen
  \bibfield  {author} {\bibinfo {author} {\bibfnamefont {Ryszard}\ \bibnamefont
  {Horodecki}}, \bibinfo {author} {\bibfnamefont {Pawe\l{}}\ \bibnamefont
  {Horodecki}}, \bibinfo {author} {\bibfnamefont {Micha\l{}}\ \bibnamefont
  {Horodecki}}, \ and\ \bibinfo {author} {\bibfnamefont {Karol}\ \bibnamefont
  {Horodecki}},\ }\bibfield  {title} {\enquote {\bibinfo {title} {Quantum
  entanglement},}\ }\href {\doibase 10.1103/RevModPhys.81.865} {\bibfield
  {journal} {\bibinfo  {journal} {Rev. Mod. Phys.}\ }\textbf {\bibinfo {volume}
  {81}},\ \bibinfo {pages} {865--942} (\bibinfo {year} {2009})}\BibitemShut
  {NoStop}%
\bibitem [{\citenamefont {Zanardi}\ \emph {et~al.}(2000)\citenamefont
  {Zanardi}, \citenamefont {Zalka},\ and\ \citenamefont
  {Faoro}}]{zanardi00entangling}%
  \BibitemOpen
  \bibfield  {author} {\bibinfo {author} {\bibfnamefont {Paolo}\ \bibnamefont
  {Zanardi}}, \bibinfo {author} {\bibfnamefont {Christof}\ \bibnamefont
  {Zalka}}, \ and\ \bibinfo {author} {\bibfnamefont {Lara}\ \bibnamefont
  {Faoro}},\ }\bibfield  {title} {\enquote {\bibinfo {title} {{Entangling power
  of quantum evolutions}},}\ }\href {\doibase 10.1103/PhysRevA.62.030301}
  {\bibfield  {journal} {\bibinfo  {journal} {Phys. Rev. A}\ }\textbf {\bibinfo
  {volume} {62}},\ \bibinfo {pages} {030301} (\bibinfo {year}
  {2000})}\BibitemShut {NoStop}%
\bibitem [{\citenamefont {{Larkin}}\ and\ \citenamefont
  {{Ovchinnikov}}(1969)}]{larkin}%
  \BibitemOpen
  \bibfield  {author} {\bibinfo {author} {\bibfnamefont {A.~I.}\ \bibnamefont
  {{Larkin}}}\ and\ \bibinfo {author} {\bibfnamefont {Yu.~N.}\ \bibnamefont
  {{Ovchinnikov}}},\ }\bibfield  {title} {\enquote {\bibinfo {title}
  {{Quasiclassical method in the theory of superconductivity}},}\ }\href
  {http://www.jetp.ac.ru/cgi-bin/e/index/e/28/6/p1200?a=list} {\bibfield
  {journal} {\bibinfo  {journal} {Soviet Journal of Experimental and
  Theoretical Physics}\ }\textbf {\bibinfo {volume} {28}},\ \bibinfo {pages}
  {1200} (\bibinfo {year} {1969})}\BibitemShut {NoStop}%
\bibitem [{\citenamefont {Maldacena}\ \emph {et~al.}(2016)\citenamefont
  {Maldacena}, \citenamefont {Shenker},\ and\ \citenamefont
  {Stanford}}]{Maldacena:2015waa}%
  \BibitemOpen
  \bibfield  {author} {\bibinfo {author} {\bibfnamefont {Juan}\ \bibnamefont
  {Maldacena}}, \bibinfo {author} {\bibfnamefont {Stephen~H.}\ \bibnamefont
  {Shenker}}, \ and\ \bibinfo {author} {\bibfnamefont {Douglas}\ \bibnamefont
  {Stanford}},\ }\bibfield  {title} {\enquote {\bibinfo {title} {{A bound on
  chaos}},}\ }\href {\doibase 10.1007/JHEP08(2016)106} {\bibfield  {journal}
  {\bibinfo  {journal} {JHEP}\ }\textbf {\bibinfo {volume} {08}},\ \bibinfo
  {pages} {106} (\bibinfo {year} {2016})},\ \Eprint
  {http://arxiv.org/abs/1503.01409} {arXiv:1503.01409 [hep-th]} \BibitemShut
  {NoStop}%
\bibitem [{\citenamefont {Lewis-Swan}\ \emph {et~al.}(2019)\citenamefont
  {Lewis-Swan}, \citenamefont {Safavi-Naini}, \citenamefont {Bollinger},\ and\
  \citenamefont {Rey}}]{rey19fotoc}%
  \BibitemOpen
  \bibfield  {author} {\bibinfo {author} {\bibfnamefont {R.~J.}\ \bibnamefont
  {Lewis-Swan}}, \bibinfo {author} {\bibfnamefont {A.}~\bibnamefont
  {Safavi-Naini}}, \bibinfo {author} {\bibfnamefont {J.~J.}\ \bibnamefont
  {Bollinger}}, \ and\ \bibinfo {author} {\bibfnamefont {A.~M.}\ \bibnamefont
  {Rey}},\ }\bibfield  {title} {\enquote {\bibinfo {title} {{Unifying
  scrambling, thermalization and entanglement through measurement of fidelity
  out-of-time-order correlators in the Dicke model}},}\ }\href {\doibase
  10.1038/s41467-019-09436-y} {\bibfield  {journal} {\bibinfo  {journal}
  {Nature Communications}\ }\textbf {\bibinfo {volume} {10}},\ \bibinfo {pages}
  {1581} (\bibinfo {year} {2019})}\BibitemShut {NoStop}%
\bibitem [{\citenamefont {Yan}\ \emph {et~al.}(2020)\citenamefont {Yan},
  \citenamefont {Cincio},\ and\ \citenamefont {Zurek}}]{yan2020loschmidt}%
  \BibitemOpen
  \bibfield  {author} {\bibinfo {author} {\bibfnamefont {Bin}\ \bibnamefont
  {Yan}}, \bibinfo {author} {\bibfnamefont {Lukasz}\ \bibnamefont {Cincio}}, \
  and\ \bibinfo {author} {\bibfnamefont {Wojciech~H.}\ \bibnamefont {Zurek}},\
  }\bibfield  {title} {\enquote {\bibinfo {title} {Information scrambling and
  loschmidt echo},}\ }\href {\doibase 10.1103/PhysRevLett.124.160603}
  {\bibfield  {journal} {\bibinfo  {journal} {Phys. Rev. Lett.}\ }\textbf
  {\bibinfo {volume} {124}},\ \bibinfo {pages} {160603} (\bibinfo {year}
  {2020})}\BibitemShut {NoStop}%
\bibitem [{Note1()}]{Note1}%
  \BibitemOpen
  \bibinfo {note} {We remark that $f$ is a function of the quantum channel
  which consists of acting with $V$ and tracing out the ancilla}\BibitemShut
  {NoStop}%
\bibitem [{\citenamefont {Collins}\ and\ \citenamefont
  {{\'S}niady}(2006)}]{collins06haar}%
  \BibitemOpen
  \bibfield  {author} {\bibinfo {author} {\bibfnamefont {Beno{\^\i}t}\
  \bibnamefont {Collins}}\ and\ \bibinfo {author} {\bibfnamefont {Piotr}\
  \bibnamefont {{\'S}niady}},\ }\bibfield  {title} {\enquote {\bibinfo {title}
  {{Integration with respect to the Haar measure on Unitary, Orthogonal and
  Symplectic group}},}\ }\href {\doibase 10.1007/s00220-006-1554-3} {\bibfield
  {journal} {\bibinfo  {journal} {Communications in Mathematical Physics}\
  }\textbf {\bibinfo {volume} {264}},\ \bibinfo {pages} {773--795} (\bibinfo
  {year} {2006})}\BibitemShut {NoStop}%
\bibitem [{\citenamefont {Casati}\ \emph {et~al.}(1979)\citenamefont {Casati},
  \citenamefont {Chirikov}, \citenamefont {Izraelev},\ and\ \citenamefont
  {Ford}}]{kicked}%
  \BibitemOpen
  \bibfield  {author} {\bibinfo {author} {\bibfnamefont {G.}~\bibnamefont
  {Casati}}, \bibinfo {author} {\bibfnamefont {B.~V.}\ \bibnamefont
  {Chirikov}}, \bibinfo {author} {\bibfnamefont {F.~M.}\ \bibnamefont
  {Izraelev}}, \ and\ \bibinfo {author} {\bibfnamefont {Joseph}\ \bibnamefont
  {Ford}},\ }\bibfield  {title} {\enquote {\bibinfo {title} {Stochastic
  behavior of a quantum pendulum under a periodic perturbation},}\ }in\
  \href@noop {} {\emph {\bibinfo {booktitle} {Stochastic Behavior in Classical
  and Quantum Hamiltonian Systems}}},\ \bibinfo {editor} {edited by\ \bibinfo
  {editor} {\bibfnamefont {Giulio}\ \bibnamefont {Casati}}\ and\ \bibinfo
  {editor} {\bibfnamefont {Joseph}\ \bibnamefont {Ford}}}\ (\bibinfo
  {publisher} {Springer Berlin Heidelberg},\ \bibinfo {address} {Berlin,
  Heidelberg},\ \bibinfo {year} {1979})\ pp.\ \bibinfo {pages}
  {334--352}\BibitemShut {NoStop}%
\bibitem [{\citenamefont {Rozenbaum}\ \emph {et~al.}(2017)\citenamefont
  {Rozenbaum}, \citenamefont {Ganeshan},\ and\ \citenamefont
  {Galitski}}]{rozenbaum17}%
  \BibitemOpen
  \bibfield  {author} {\bibinfo {author} {\bibfnamefont {Efim~B.}\ \bibnamefont
  {Rozenbaum}}, \bibinfo {author} {\bibfnamefont {Sriram}\ \bibnamefont
  {Ganeshan}}, \ and\ \bibinfo {author} {\bibfnamefont {Victor}\ \bibnamefont
  {Galitski}},\ }\bibfield  {title} {\enquote {\bibinfo {title} {{Lyapunov
  exponent and out-of-time-ordered correlator's growth rate in a chaotic
  system}},}\ }\href {\doibase 10.1103/PhysRevLett.118.086801} {\bibfield
  {journal} {\bibinfo  {journal} {Phys. Rev. Lett.}\ }\textbf {\bibinfo
  {volume} {118}},\ \bibinfo {pages} {086801} (\bibinfo {year}
  {2017})}\BibitemShut {NoStop}%
\bibitem [{\citenamefont {Xu}\ \emph {et~al.}(2020)\citenamefont {Xu},
  \citenamefont {Scaffidi},\ and\ \citenamefont {Cao}}]{xu2020}%
  \BibitemOpen
  \bibfield  {author} {\bibinfo {author} {\bibfnamefont {Tianrui}\ \bibnamefont
  {Xu}}, \bibinfo {author} {\bibfnamefont {Thomas}\ \bibnamefont {Scaffidi}}, \
  and\ \bibinfo {author} {\bibfnamefont {Xiangyu}\ \bibnamefont {Cao}},\
  }\bibfield  {title} {\enquote {\bibinfo {title} {{Does scrambling equal
  chaos?}}}\ }\href {\doibase 10.1103/PhysRevLett.124.140602} {\bibfield
  {journal} {\bibinfo  {journal} {Phys. Rev. Lett.}\ }\textbf {\bibinfo
  {volume} {124}},\ \bibinfo {pages} {140602} (\bibinfo {year}
  {2020})}\BibitemShut {NoStop}%
\bibitem [{\citenamefont {Lipkin}\ \emph {et~al.}(1965)\citenamefont {Lipkin},
  \citenamefont {Meshkov},\ and\ \citenamefont {Glick}}]{LIPKIN1965188}%
  \BibitemOpen
  \bibfield  {author} {\bibinfo {author} {\bibfnamefont {H.J.}\ \bibnamefont
  {Lipkin}}, \bibinfo {author} {\bibfnamefont {N.}~\bibnamefont {Meshkov}}, \
  and\ \bibinfo {author} {\bibfnamefont {A.J.}\ \bibnamefont {Glick}},\
  }\bibfield  {title} {\enquote {\bibinfo {title} {Validity of many-body
  approximation methods for a solvable model: (i). exact solutions and
  perturbation theory},}\ }\href {\doibase
  https://doi.org/10.1016/0029-5582(65)90862-X} {\bibfield  {journal} {\bibinfo
   {journal} {Nuclear Physics}\ }\textbf {\bibinfo {volume} {62}},\ \bibinfo
  {pages} {188 -- 198} (\bibinfo {year} {1965})}\BibitemShut {NoStop}%
\bibitem [{\citenamefont {Glick}\ \emph {et~al.}(1965)\citenamefont {Glick},
  \citenamefont {Lipkin},\ and\ \citenamefont {Meshkov}}]{GLICK1965211}%
  \BibitemOpen
  \bibfield  {author} {\bibinfo {author} {\bibfnamefont {A.J.}\ \bibnamefont
  {Glick}}, \bibinfo {author} {\bibfnamefont {H.J.}\ \bibnamefont {Lipkin}}, \
  and\ \bibinfo {author} {\bibfnamefont {N.}~\bibnamefont {Meshkov}},\
  }\bibfield  {title} {\enquote {\bibinfo {title} {Validity of many-body
  approximation methods for a solvable model: (iii). diagram summations},}\
  }\href {\doibase https://doi.org/10.1016/0029-5582(65)90864-3} {\bibfield
  {journal} {\bibinfo  {journal} {Nuclear Physics}\ }\textbf {\bibinfo {volume}
  {62}},\ \bibinfo {pages} {211 -- 224} (\bibinfo {year} {1965})}\BibitemShut
  {NoStop}%
\bibitem [{\citenamefont {Meshkov}\ \emph {et~al.}(1965)\citenamefont
  {Meshkov}, \citenamefont {Glick},\ and\ \citenamefont
  {Lipkin}}]{MESHKOV1965199}%
  \BibitemOpen
  \bibfield  {author} {\bibinfo {author} {\bibfnamefont {N.}~\bibnamefont
  {Meshkov}}, \bibinfo {author} {\bibfnamefont {A.J.}\ \bibnamefont {Glick}}, \
  and\ \bibinfo {author} {\bibfnamefont {H.J.}\ \bibnamefont {Lipkin}},\
  }\bibfield  {title} {\enquote {\bibinfo {title} {Validity of many-body
  approximation methods for a solvable model: (ii). linearization
  procedures},}\ }\href {\doibase https://doi.org/10.1016/0029-5582(65)90863-1}
  {\bibfield  {journal} {\bibinfo  {journal} {Nuclear Physics}\ }\textbf
  {\bibinfo {volume} {62}},\ \bibinfo {pages} {199 -- 210} (\bibinfo {year}
  {1965})}\BibitemShut {NoStop}%
\bibitem [{Note2()}]{Note2}%
  \BibitemOpen
  \bibinfo {note} {See Supplemental Material at [URL will be inserted by
  publisher], which contains Ref.~\cite {papparlardi18,xu2020}, for numerical
  results on the Lipkin-Meshkov-Glick model.}\BibitemShut {Stop}%
\bibitem [{\citenamefont {Li}\ \emph {et~al.}(2018)\citenamefont {Li},
  \citenamefont {Chen},\ and\ \citenamefont {Fisher}}]{li18transition}%
  \BibitemOpen
  \bibfield  {author} {\bibinfo {author} {\bibfnamefont {Yaodong}\ \bibnamefont
  {Li}}, \bibinfo {author} {\bibfnamefont {Xiao}\ \bibnamefont {Chen}}, \ and\
  \bibinfo {author} {\bibfnamefont {Matthew P.~A.}\ \bibnamefont {Fisher}},\
  }\bibfield  {title} {\enquote {\bibinfo {title} {{Quantum Zeno effect and the
  many-body entanglement transition}},}\ }\href {\doibase
  10.1103/PhysRevB.98.205136} {\bibfield  {journal} {\bibinfo  {journal} {Phys.
  Rev. B}\ }\textbf {\bibinfo {volume} {98}},\ \bibinfo {pages} {205136}
  (\bibinfo {year} {2018})}\BibitemShut {NoStop}%
\bibitem [{\citenamefont {Skinner}\ \emph {et~al.}(2019)\citenamefont
  {Skinner}, \citenamefont {Ruhman},\ and\ \citenamefont
  {Nahum}}]{nahum18transition}%
  \BibitemOpen
  \bibfield  {author} {\bibinfo {author} {\bibfnamefont {Brian}\ \bibnamefont
  {Skinner}}, \bibinfo {author} {\bibfnamefont {Jonathan}\ \bibnamefont
  {Ruhman}}, \ and\ \bibinfo {author} {\bibfnamefont {Adam}\ \bibnamefont
  {Nahum}},\ }\bibfield  {title} {\enquote {\bibinfo {title}
  {{Measurement-Induced Phase Transitions in the Dynamics of Entanglement}},}\
  }\href {\doibase 10.1103/PhysRevX.9.031009} {\bibfield  {journal} {\bibinfo
  {journal} {Phys. Rev. X}\ }\textbf {\bibinfo {volume} {9}},\ \bibinfo {pages}
  {031009} (\bibinfo {year} {2019})}\BibitemShut {NoStop}%
\bibitem [{\citenamefont {Bao}\ \emph {et~al.}(2020)\citenamefont {Bao},
  \citenamefont {Choi},\ and\ \citenamefont {Altman}}]{bao19theory}%
  \BibitemOpen
  \bibfield  {author} {\bibinfo {author} {\bibfnamefont {Yimu}\ \bibnamefont
  {Bao}}, \bibinfo {author} {\bibfnamefont {Soonwon}\ \bibnamefont {Choi}}, \
  and\ \bibinfo {author} {\bibfnamefont {Ehud}\ \bibnamefont {Altman}},\
  }\bibfield  {title} {\enquote {\bibinfo {title} {{Theory of the phase
  transition in random unitary circuits with measurements}},}\ }\href {\doibase
  10.1103/PhysRevB.101.104301} {\bibfield  {journal} {\bibinfo  {journal}
  {Phys. Rev. B}\ }\textbf {\bibinfo {volume} {101}},\ \bibinfo {pages}
  {104301} (\bibinfo {year} {2020})}\BibitemShut {NoStop}%
\bibitem [{\citenamefont {Jian}\ \emph {et~al.}(2020)\citenamefont {Jian},
  \citenamefont {You}, \citenamefont {Vasseur},\ and\ \citenamefont
  {Ludwig}}]{jian19theory}%
  \BibitemOpen
  \bibfield  {author} {\bibinfo {author} {\bibfnamefont {Chao-Ming}\
  \bibnamefont {Jian}}, \bibinfo {author} {\bibfnamefont {Yi-Zhuang}\
  \bibnamefont {You}}, \bibinfo {author} {\bibfnamefont {Romain}\ \bibnamefont
  {Vasseur}}, \ and\ \bibinfo {author} {\bibfnamefont {Andreas W.~W.}\
  \bibnamefont {Ludwig}},\ }\bibfield  {title} {\enquote {\bibinfo {title}
  {{Measurement-induced criticality in random quantum circuits}},}\ }\href
  {\doibase 10.1103/PhysRevB.101.104302} {\bibfield  {journal} {\bibinfo
  {journal} {Phys. Rev. B}\ }\textbf {\bibinfo {volume} {101}},\ \bibinfo
  {pages} {104302} (\bibinfo {year} {2020})}\BibitemShut {NoStop}%
\bibitem [{\citenamefont {Pappalardi}\ \emph {et~al.}(2018)\citenamefont
  {Pappalardi}, \citenamefont {Russomanno}, \citenamefont {Zunkovic},
  \citenamefont {Iemini}, \citenamefont {Silva},\ and\ \citenamefont
  {Fazio}}]{papparlardi18}%
  \BibitemOpen
  \bibfield  {author} {\bibinfo {author} {\bibfnamefont {Silvia}\ \bibnamefont
  {Pappalardi}}, \bibinfo {author} {\bibfnamefont {Angelo}\ \bibnamefont
  {Russomanno}}, \bibinfo {author} {\bibfnamefont {Bojan}\ \bibnamefont
  {Zunkovic}}, \bibinfo {author} {\bibfnamefont {Fernando}\ \bibnamefont
  {Iemini}}, \bibinfo {author} {\bibfnamefont {Alessandro}\ \bibnamefont
  {Silva}}, \ and\ \bibinfo {author} {\bibfnamefont {Rosario}\ \bibnamefont
  {Fazio}},\ }\bibfield  {title} {\enquote {\bibinfo {title} {Scrambling and
  entanglement spreading in long-range spin chains},}\ }\href {\doibase
  10.1103/PhysRevB.98.134303} {\bibfield  {journal} {\bibinfo  {journal} {Phys.
  Rev. B}\ }\textbf {\bibinfo {volume} {98}},\ \bibinfo {pages} {134303}
  (\bibinfo {year} {2018})}\BibitemShut {NoStop}%
\end{thebibliography}%
	
\pagebreak
\begin{widetext}

\begin{center}
	{\bf\Large Supplemental Material} 
\end{center}

	In this Supplemental Material, we present numerical results on the recovery ratio of a qubit coupled to a semiclassical Lipkin-Meshkov-Glick model. We define the latter by the Hamiltonian
	\begin{equation}
	    H_{\text{LMG}} = X_c + 2 Z_c^2 / S 
	\end{equation}
	where $X_c, Y_c, Z_c$ are SU(2) spin operators with a large spin $S$ in the semiclassical limit (the effective Planckian constant $\hbar \sim 1/S$). 	We recall that this model is not chaotic. Yet, out-of-time order correlators can grow exponentially due to saddle-dominated scrambling~\cite{papparlardi18,xu2020}. Meanwhile, we denote by $X_q, Y_q, Z_q$ the spin-half operators acting on the qubit. The qubit is coupled to the LMG model as 
	\begin{equation}
	    H_{cq} = h X_q + J Z_q Z_c / S \,. 
	\end{equation}
	In both the above equations, the factor $1/S$ ensures a correct classical limit as $S \to \infty$. The total Hamiltonian is thus
	\begin{equation}
	    H = H_{cq} + H_{\text{LMG}} \,.
	\end{equation}
	Following the protocol in the main text, we prepare the LMG model at infinite temperature (maximally mixed state), and the qubit in a pure state. The intruder's action is a strong measurement in the $Z_q$ basis. Some numerical results of the recovery ratio are shown in Fig.~\ref{fig:supp}. We find that for almost any initial qubit state, there is an extended period during which the recovery ratio is close to the chaotic-classical-bath prediction $r_{\text{c}} = 1/3$ (the crossover to the fully quantum regime where $r_{\text{q}} = 1/2$ is much slower than in presence of classical chaos). Moreover, further numerics shows that the classical plateau does not depend on the saddle-dominated scrambling. In fact, we believe that it is simply due to the mixed initial state of the bath: if one initializes the bath in a semiclassical (coherent) state, that will strongly affect the recovery ratio, which is no longer approximated by $ r_{\text{c}} = 1/3$. Therefore, the classical plateau without chaos is less robust than the chaotic counterpart studied in the main text. 
		\begin{figure*}
	    \centering
	    \includegraphics[scale=.6]{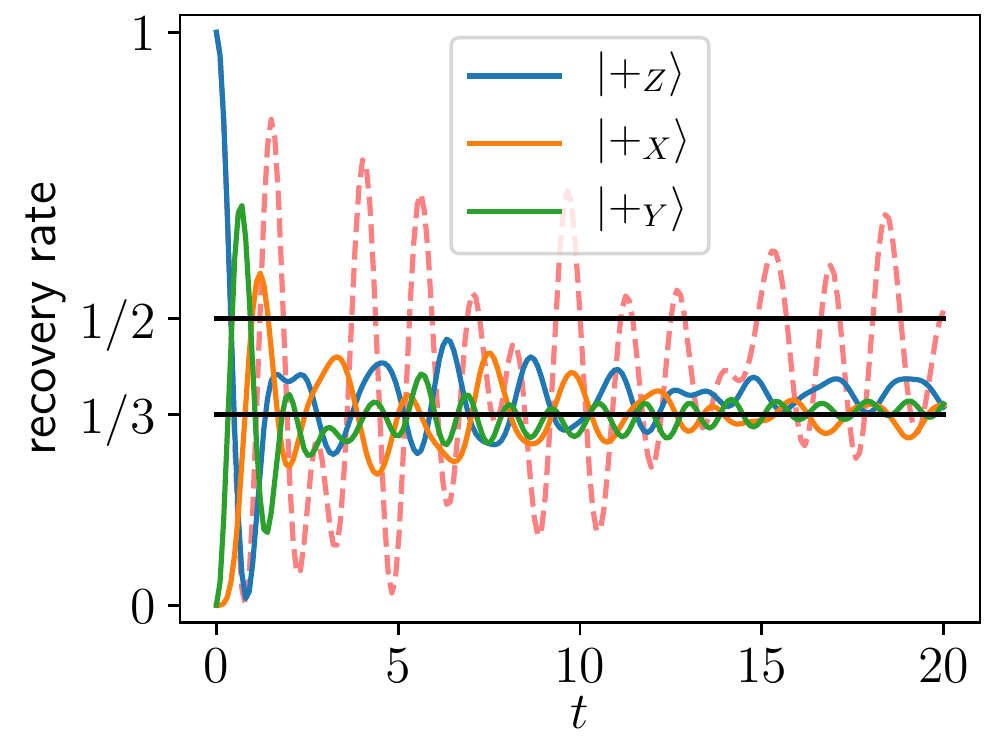}
	    \includegraphics[scale=.6]{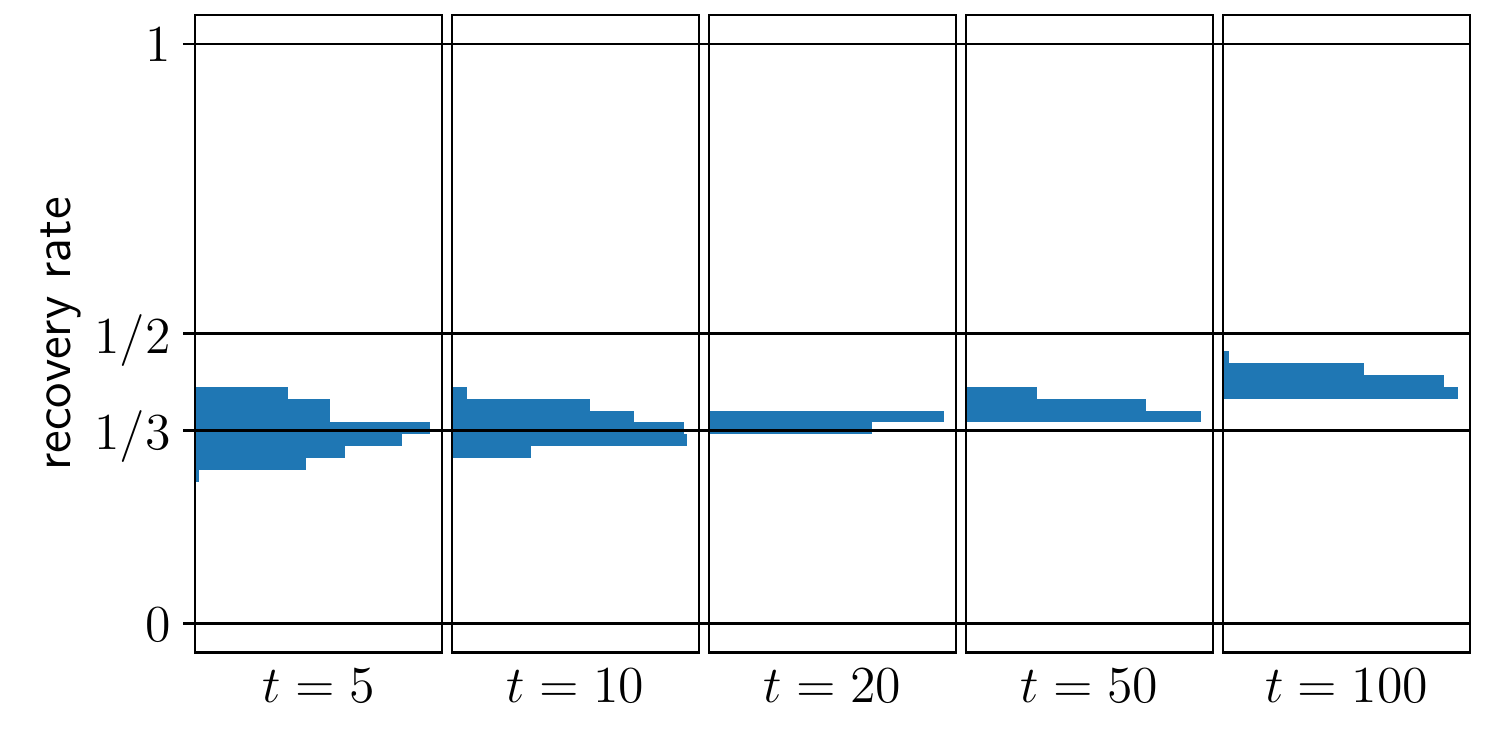}
	    \caption{Recovery ratio in a model of a qubit coupled to a semiclassical LMG model with spin $S=100$, $J = 2.4$ and $h = 2$.  \textit{Left}. Solid curves: Recovery ratio as function of time for three initial qubit states, when the bath is initialized to be maximally mixed. Dashed curve: the recovery ratio when the bath is initially a coherent (semiclassical) state with $(X_c, Y_c, Z_c) \propto (1,0,0)$ [the qubit's initial state is $\vert +_Z \rangle$]. We observe significant oscillation instead of a plateau. \textit{Right}. Distribution of recovery ratio at $t = 5, 10,20, 50, 100$ when the qubit initial state is randomly and uniformly chosen on the Bloch sphere. $300$ samples are calculated for each histogram.}
	    \label{fig:supp}
	\end{figure*}
\end{widetext}
	\end{document}